\documentclass[prb,twocolumn,superscriptaddress]{revtex4-1}
\usepackage{amsmath}
\usepackage{color}
\usepackage{graphicx}
\usepackage{dcolumn}
\usepackage{bm}
\usepackage[dvipdfm,bookmarks=true,colorlinks,%
            citecolor=blue,linkcolor=blue, hypertex, %
            breaklinks=true]{hyperref}
\usepackage{setspace}
\usepackage{subfigure}
\usepackage{enumerate}

\begin{document}

\title{Degenerate orbital effect in a three-orbital periodic Anderson model}
\author{Jianwei Yang}
\affiliation{Department of Physics, Beijing Normal University, Beijing 100875, China}
\affiliation{Beijing Computational Science Research Center, Beijing 100193, China}
\author{Qiyu Wang}
\affiliation{Department of Physics, Beijing Normal University, Beijing 100875, China}
\author{Tianxing Ma}
\affiliation{Department of Physics, Beijing Normal University, Beijing 100875, China}
\affiliation{Beijing Computational Science Research Center, Beijing 100193, China}
\author{Qiaoni Chen}
\email{qiaoni@bnu.edu.cn}
\affiliation{Department of Physics, Beijing Normal University, Beijing 100875, China}

\begin{abstract}
The competition between the Ruderman-Kittel-Kasuya-Yosida effect and Kondo effect is a central subject of
 the periodic Anderson model. By using the density matrix embedding theory, we study a three-orbital periodic
 Anderson model, in which the effects of degenerate conduction orbitals, via the local magnetic moments, number
 of electrons, and spin-spin correlation functions, are investigated. From the phase diagram at half filling, we
 find there exist two different antiferromagnetic phases and one paramagnetic phase. To explore the difference
 between the two antiferromagnetic phases, the topology of the Fermi surface and the connection with the standard
 periodic Anderson model are considered. The spin-spin correlation functions yield insight into the competition
 between Ruderman-Kittel-Kasuya-Yosida interaction and Kondo interaction. We further find there exist "scaling
 transformations," and by applying them to the data with different hybridization strength, all the data collapses.
Our calculations agree with previous studies on the standard periodic Anderson model.

\end{abstract}

\maketitle

\section{Introduction}
The accessible of clean interfaces between transition metal oxides provides new opportunities for electronics \cite{Mannhart2010_Science327-1607-1611,Charlebois2013_PRB87-035137}. The reason is that most of the traditional electronic devices are fabricated with semiconductor materials, whose behaviors are more predictable since the electron-electron interactions are not dominant. Both transition metal oxides and rare earth compounds are considered as strongly correlated material, since the transition metal oxides include elements which have partially filled $d$ orbitals, while lanthanides and actinides compounds have partially filled $f$ orbitals. Successful fabrications of the layered superlattices of heavy fermion material \cite {Shishido2010_Science327-980--983,Mizukami2011_NatPhys7-849,Goh2012_PRL109-157006} made a step towards strongly correlated electronic devices, and triggered many interesting studies on layered $f$-electron systems
 \cite{Peters2013_PRB88-155134,Tada2013_PRB88-235121,Peters2014_PRB89-041106,Sen2015_PRB91-155146,Sen2016_PRB93-155136, Hu2017_PRB95-235122}. However several fundamental problems still exist, in both theoretical and computational aspects.

One of the fascinating questions is what occurs at the interface of normal metal and strongly correlated insulator. The answer is Kondo proximity effect by dynamical mean field theory (DMFT) \cite{Helmes2008_PRL101-066802}, and Kondo screening embraces both sides of the interface by determinant quantum Monte carlo (DQMC) \cite{Euverte2012_PRL108-246401}.  Two neighboring conduction electron layers and one localized electron layer is considered  to describe this problem \cite{Sen2016_PRB93-155136, Hu2017_PRB95-235122}. Meanwhile the correlated layers sandwiched between normal metallic layers \cite{Zenia2009_PRL103-116402} and an even more complex structure \cite{Zujev2013_PRB88-094415} is another interesting problem. In order to understand more about this problem, we start with a quasi-two dimensional model. The model includes three layers, and the electrons in the correlated layer is allowed to hop to the other two conduction layers.

In the context of periodic Anderson model (PAM), the model we studied could be interpreted as two orthogonal conduction orbitals and one localized orbital on each site. Therefore besides the connection with layered $f$-electron system, our work are related with the traditional degenerate orbitals problem in the fields of heavy fermions. It was pointed out that multi-orbital effect plays an essential role in some uranium-based compounds \cite{Cox1987_PRL59-1240--1243}, and PAM which includes degenerated $f$ orbitals has been studied by DMFT \cite{Koga2003_JPCM15-S2215}. Moreover it was suggested that multi-orbital conduction electrons may be relevant to the heavy-fermion behavior of 3d transitional metal compounds LiV$_2$O$_4$ \cite{Yamashita2003_PRB67-195107}. Apart from the degenerate orbital effect, the model is also connected with the multichannel Kondo problem. Various mechanisms of non-Fermi-liquid behavior were discussed based on a multichannel Kondo lattice model \cite{Irkhin2016_EPJB89-117}. Compared with the Kondo lattice model, the PAM includes charge degree of freedom, so the physics in it would be more rich.

By employing the density matrix embedding theory (DMET)  \cite{Knizia2012_PRL109-186404,Knizia2013_JCTC9-1428-1432} , we study a three orbital PAM in this paper. We calculate local magnetic moments, number of electrons, and spin-spin correlation functions to understand the physics in the model. In the following we will describe the model first, give a brief introduction to the method, present our results in detail, and in the end we make a summary and conclusion.

\section{Model and Methods}
We consider a three orbital PAM on a two dimensional square lattice, the Hamiltonian is the following
\begin{equation}
\begin{aligned}
H=&-t\sum_{<ij>\sigma} (c_{i\sigma}^\dag c_{j\sigma}+h.c)-t\sum_{<ij>\sigma} (d_{i\sigma}^\dag d_{j\sigma}+h.c)\\&+V_1\sum_{i\sigma} (c_{i\sigma}^\dag f_{i\sigma}+h.c)+V_2\sum_{i\sigma} (d_{i\sigma}^\dag f_{i\sigma}+h.c)\\&+E_f \sum_{i\sigma} f_{i\sigma}^\dag f_{i\sigma}+U\sum_i n_{i\uparrow}^f n_{i\downarrow}^f
\end{aligned}
\end{equation}
where $c_{i\sigma}^\dag(c_{i\sigma})$ and  $d_{i\sigma}^\dag(d_{i\sigma})$ are the creation (annihilation) operators of the two conduction orbitals on site $i$ with spin $\sigma$, and $f_{i\sigma}^\dag(f_{i\sigma})$ is the creation (annihilation) operators of localized orbital.
$t$ is the hopping integral between nearest-neighboring conduction orbitals, $E_f$ is the onsite energy of the localized orbital ($f$ state), which defines the relative position of $f$ state with respect to the Fermi energy of the conduction orbital ($c$ state) . $V_1$($V_2$) is the hybridization strength between $c$($d$) and $f$ states on the same site, and $U$ is the on-site Coulomb repulsion of the $f$ states.

\begin{figure}
\centering
\includegraphics[width=0.225\textwidth]{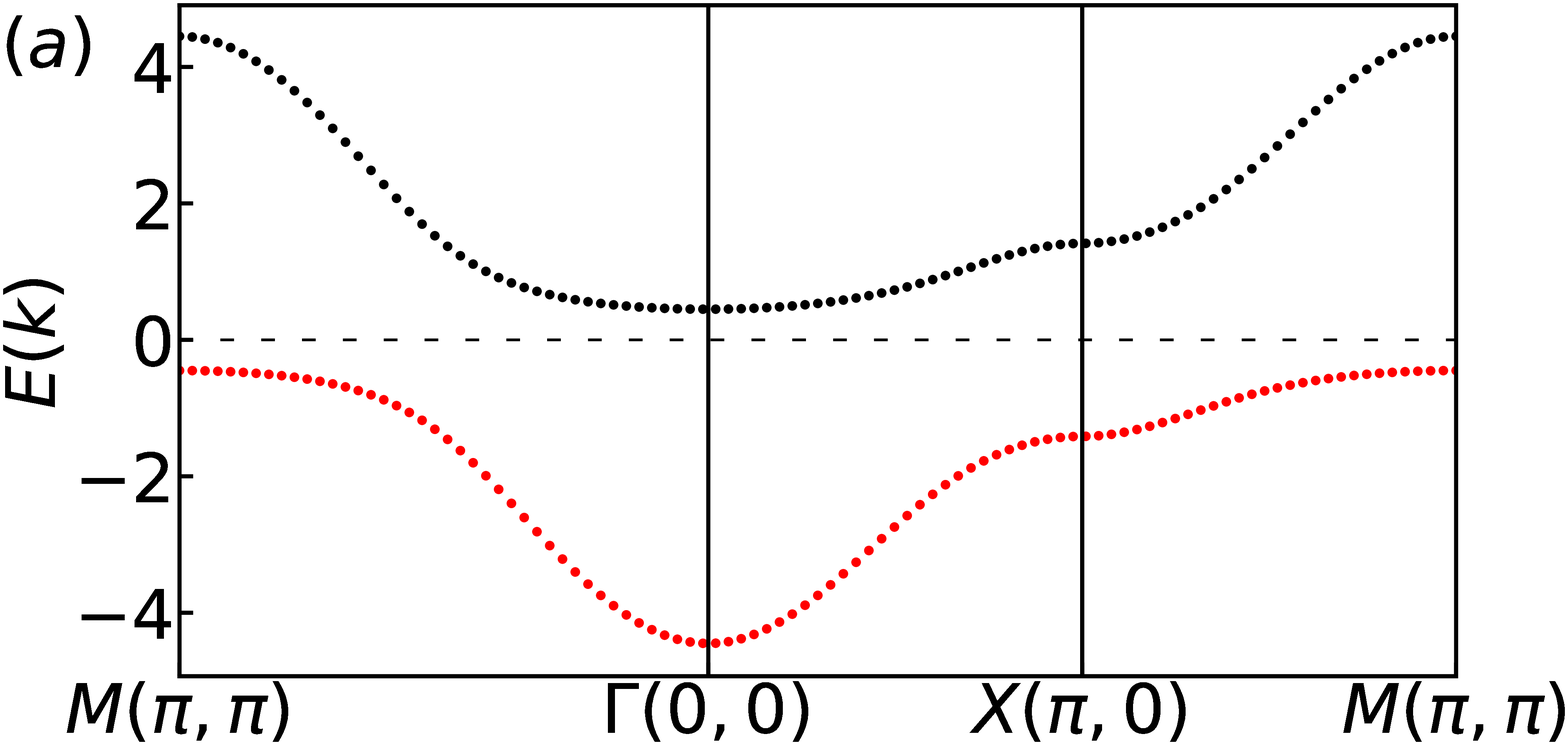}
\includegraphics[width=0.225\textwidth]{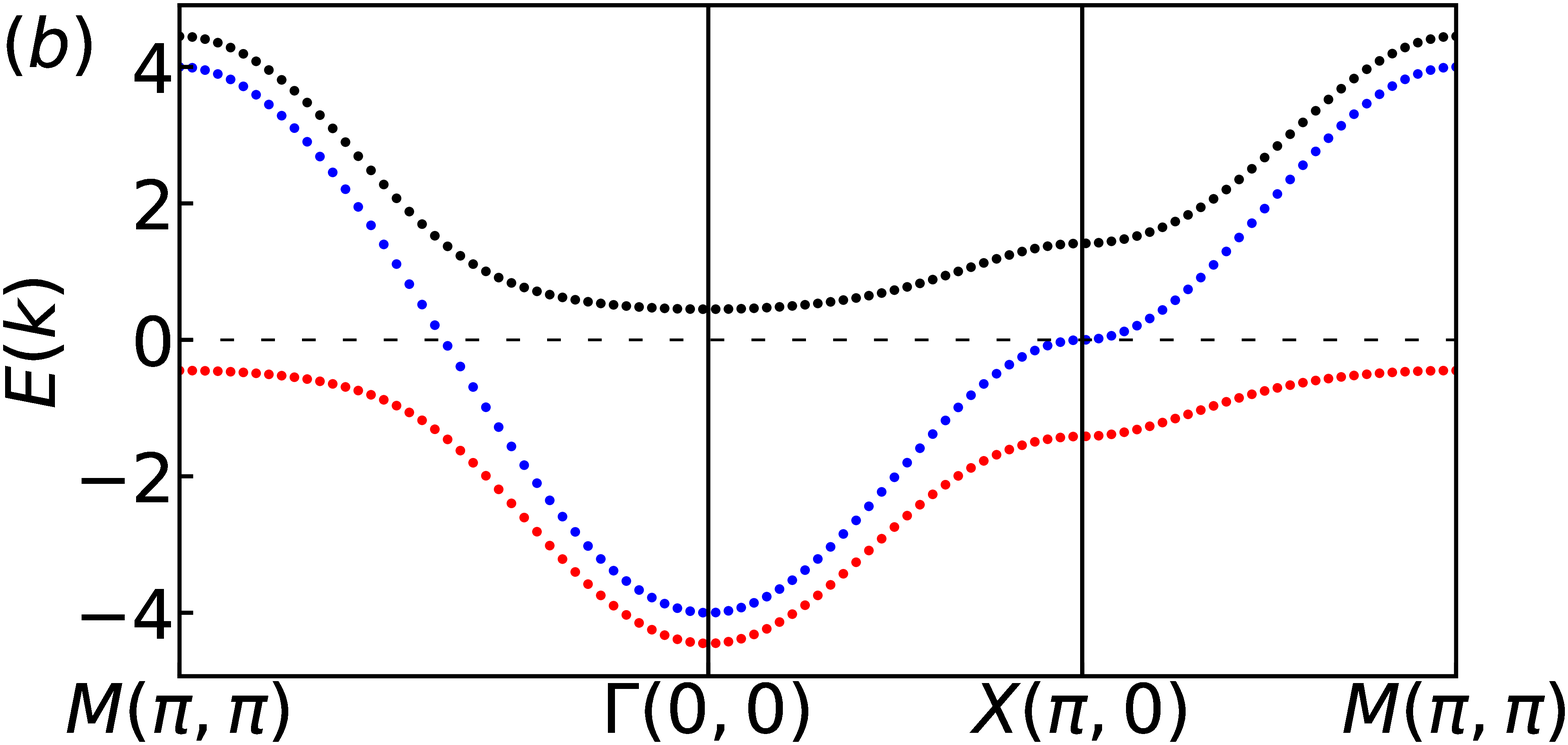}
\caption{(Color online) The non-interacting dispersion relations on the square lattice when $E_f=0$ and $V_1=V_2=V=1.0t$. (a) Standard PAM which has one conduction orbital and one localized orbital on each site, the hybridization between the two orbitals open a gap. (b) Three orbital PAM, which has two conduction orbitals and one localized orbital on each site. The shape of the upper band (black line) and the lower band (red line) is same as the standard PAM.  The black dashed line is the Fermi level at half filling.}
\label{Dispersion}
\end{figure}

In the non-interacting case, the Hamiltonian in the momentum space could be written as:
\begin{equation}
H_0=\sum_{\mathbf{k},\sigma}
    \left(\begin{array}{ccc}
    c^\dag_{\mathbf{k}\sigma} & d^\dag_{\mathbf{k}\sigma} & f^\dag_{\mathbf{k}\sigma }
    \end{array}\right)
    \left(\begin{array}{ccc}
    \epsilon_{\mathbf{k}} & 0 & V_1\\
    0 & \epsilon_{\mathbf{k}}  & V_2 \\
    V_1 & V_2 & E_f
  \end{array}\right)
  \left(\begin{array}{c}
   c_{\mathbf{k}\sigma }\\ d_{\mathbf{k}\sigma} \\ f_{\mathbf{k}\sigma}
    \end{array}\right)
\end{equation}
here $\epsilon(\mathbf{k})=-2t(cosk_x+cosk_y)$ is the dispersion relation of the conduction band. Diagonalizing the non-interacting Hamiltonian $H_0$ yields three bands:
\begin{equation}
E(\mathbf{k})=
\begin{cases}
\frac{1}{2}\left[ E_f + \epsilon(\mathbf{k}) + \sqrt{\left(E_f-\epsilon(\mathbf{k})\right)^2 + 4V_1^2+4V_2^2} \right] \\
\epsilon(\mathbf{k})\\
\frac{1}{2}\left[ E_f + \epsilon(\mathbf{k}) - \sqrt{\left(E_f-\epsilon(\mathbf{k})\right)^2 + 4V_1^2+4V_2^2} \right] \\
\end{cases}
\end{equation}

We plot the dispersion relation in Fig.\ref{Dispersion} . As comparison the dispersion of the standard PAM is shown in the left panel. The hybridization between the conduction orbital and localized orbital results in two different bands, and produces a gap between the two bands (black and red). It is not difficult to prove that the gap always exists no matter how the parameters change. The dispersion relation of the three orbital PAM is similar. Except there is an additional band in the middle the other two bands. The additional band is shown as the blue line in the right panel of Fig. \ref{Dispersion}. It can be proved that the blue band is always in the middle of the black and red bands. The shape of the black band and the red band is similar to the ordinary PAM. The dispersion of the additional band is the same as the conduction band, because it is a linear combination of the two conduction orbitals. At half filling the ordinary PAM is insulating, while the three orbital PAM is metallic.

Ever since it was developed DMET \cite{Knizia2012_PRL109-186404,Knizia2013_JCTC9-1428-1432} has been applied to several different areas. Including standard Hubbard model \cite{Chen2014_PRB89-165134}, Hubbard-Holestein model\cite{Sandhoefer2016_PRB94-085115} which contains electron-phonon interaction, cuprates\cite{Zheng2016_PRB93-035126,Zheng2017_Science358-1155--1160}, single impurity Anderson model\cite{Mukherjee2017_Phys.Rev.B95-155111}, as well as quantum molecules\cite{Sun2014_JCTC10-3784-3790,Wouters2016_JCTC12-2706-2719}. Besides ground-state static properties, dynamic properties such as spectral function \cite{Booth2015_PRB91-155107} could be derived, and so does the non-equilibrium dynamics \cite{Kretchmer2018_JChemPhys148-054108}. For more details of the methods, please refer to the thesis, Ref.\cite{Zheng2017_PU}.

\begin{figure}
\centering
\includegraphics[width=0.35\textwidth]{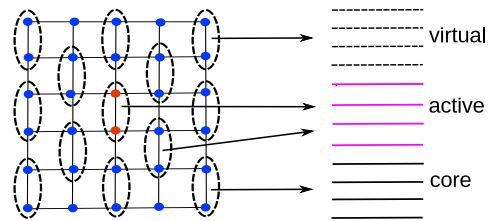}
\caption{(Color online) In the DMET the square lattice is first divided into clusters ($1\times2$ here), and one of the cluster is chosen as impurity sites (the red sites), the rest blue sites of the lattice is considered as environment sites. The bath orbitals, core orbitals and virtual orbitals are linear combination of orbitals environment sites. The impurity orbitals and bath orbitals constitute active space.}
\label{Cluster}
\end{figure}

In a DMET calculation, the lattice sites are first divided into different clusters as shown in Fig.\ref{Cluster}. The clusters are chosen to tile the whole lattice, and they are always the unitcells of lattice in order to keep the translation invariance. The example of $1 \times 2$ cluster are displayed in Fig. \ref{Cluster}.

An auxiliary system with Hamiltonian $h$ is then introduced:
\begin{equation}
\begin{aligned}
h=&h_0+v
\end{aligned}
\end{equation}
where $h_0$ is the one body terms in $H$, and $v$ is the correlation potential within cluster. In the particle number conserving case (no superconducting phase) $v$ has the form:
\begin{equation}
\begin{aligned}
v=&\sum_{C}\sum_{i,j\in C}v_{ij\sigma}c_{i\sigma}^\dag c_{j\sigma}
\end{aligned}
\end{equation}
here $C$ is one of the clusters that within dashed circles in Fig.\ref{Cluster}. $h$ is block diagonal since $v$ is only within cluster, and $v$ is a replacement of local interaction.

The one-body Hamiltonian $h$ is simple enough to be solved. From the ground state $|\Phi\rangle$ of $h$, the embedding basis could be constructed. The sites in one of the cluster (the red sites in Fig.\ref{Cluster}) are chosen as the impurity orbitals. The remaining sites (the blue sites in Fig.\ref{Cluster}) are the environment orbitals. There are several mathematically equivalent unitary transformations after apply which the environment orbitals are linearly combined into bath orbitals (magenta energy levels in Fig.\ref{Cluster}), core orbitals (black energy levels) and virtual orbitals (grey energy levels). Core (virtual) orbitals are completely full (empty), thus only bath orbitals are entangled with impurity orbitals. The core orbitals and bath orbitals constitute active space, and the number of bath orbitals is at most the number of impurity orbitals. The impurity Hamiltonian $H_{imp}$ is constructed as:
\begin{equation}
\begin{aligned}
H_{imp} = PhP -\sum_{i,j\in imp}v_{ij}c_{i\sigma}^\dag c_{j\sigma}
+ U\sum_{i\in imp}n_{i\uparrow}^f n_{i\downarrow}^f
\end{aligned}
\end{equation}
here $P$ is the projection operator which projects the system to the active space. The correlation potentials on the impurity orbitals are replaced by the onsite Coulomb interaction. Since the impurity Hamiltonian $H_{imp}$ is only within the active space, Exact diagonalization and other computational expensive methods could be used to solve the ground state $|\Psi\rangle$ of $H_{imp}$. In this work we use density matrix renormalization group (DMRG) to solve the impurity model $H_{imp}$.

The impurity model $H_{imp}$ include a few impurity orbitals as well as a few bath orbitals. The one-body terms in $H_{imp}$ are of a general form, so the real space DMRG is not suitable for this problem. Instead momentum space DMRG which are widely used in quantum chemistry simulations is appropriate. Our simulations are finished with the BLOCK quantum chemistry DMRG package \cite{Sharma2012_J.Chem.Phys.136-124121}. Since the cluster is chosen as $1\times2$, there are $6$ impurity orbitals and $6$ bath orbitals in the impurity model. Thus in a DMRG calculation the impurity model has $12$ orbitals, and mostly $12$ electrons. The precision and computational cost of a DMRG calculation depends on the number of states kept $M$. In most of our simulations $M=3000$ is enough, but near phase transition $M=10000$ is required.

The corresponding 1 particle reduced density matrix (1-PDM) of the ground state $|\Psi\rangle$ of $H_{imp}$ is $\rho^{I}$, and the correlation potential is updated through $\rho^{I}$. Our goal is to minimize the difference between $|\Psi\rangle$ and $|\Phi\rangle$ (ground state of $h$). This is accomplished by first downfolding $|\Phi\rangle$ to the active space $|\phi\rangle=P|\Phi\rangle$, and the 1-PDM of $|\phi\rangle$ is $\rho^{0}$. Both $\rho^{I}$ and $\rho^{0}$ are dependent on the correlation potential $v$. However $\rho^{I}$ is much more computational costly than $\rho^{0}$. In the process to update the new correlation potential, $\rho^{I}$ is fixed and only $\rho^{0}$ is changed with the correlation potential $v$.     

\begin{equation}
\begin{aligned}
\min \limits_{v} f(v)=\sqrt{\sum_{ij}|\rho^{I}_{ij}(v_0)-\rho^{0}_{ij}(v)|^2}
\end{aligned}
\end{equation}

When the optimal $v$ is found, it's used to update the auxiliary Hamiltonian $h$ and its ground state $|\Phi\rangle$, the embedding basis, the impurity model $H_{imp}$, as well as the corresponding $|\Psi\rangle$ and $\rho^{I}$. Thus the self-consistent loop is formed.


In summary the DMET calculations proceed the following steps:
\begin{enumerate}[(1)]
\item An initial guess of the correlation potential $v_{0}$ is chosen.
\item Solve the auxiliary lattice Hamiltonian to obtain the lattice wave function $|\Phi\rangle$.
\item Embedding basis is constructed from the lattice wave function $|\Phi\rangle$.
\item Transform to the embedding basis, and add the interaction to get impurity model $H_{imp}$.
\item Using DMRG impurity solver to compute the ground state $|\Psi\rangle$ of the impurity model, and calculate the corresponding 1-PDM $\rho^{I}$.
\item Update the correlation potential $v$ to minimize the difference of $\rho^{I}$ and $\rho^{0}$ .
\item Go back to step (2) until the correlation potential $v$ converges.
\end{enumerate}

The local observables such as local magnetic moment and the number of electrons are extracted directly from 1-PDM of $|\Phi\rangle$. Other observables such as ground state energy and spin-spin correlation are calculated from 2-PDM of $|\Phi\rangle$.

\section{Results}
We have run the DMET calculations of the three orbital PAM on a two dimensional square lattice. The lattice size in our calculation is $200\times200$. We mainly focus on the physics at half filling, and in our simulation $t=1$, $U=8$.

\subsection{Order parameter and phase diagram}
\begin{figure}
\includegraphics[height=0.3\textwidth,width=0.5\textwidth]{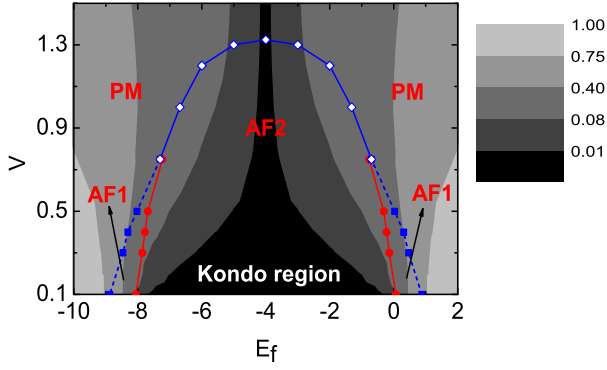}
\centering
\caption{(Color online) Ground states phase diagram when $U=8t$, it is symmetric with respect to $E_f=-U/2$. There are three different phases, para-magnetic (PM) phase and two anti-ferromagnetic phases (AF1 and AF2). The phase transitions from AF2 phase to the other two phases are first order,  and marked as red solid line and blue solid line. The phase transition between AF1 phase and PM phase are continuous, and marked as blue dashed line.  Within the AF2 phase there's a Kondo region, in which $n_f\approx1.0$. The grey scale is $x=|n_f-1.0|$.  }
\label{PD}
\end{figure}

First we focus on the symmetric case when $V_1=V_2=V$. The ground state phase diagram at half filling is shown in Fig. \ref{PD}, and it's symmetric with respect to $E_f=-\frac{U}{2}$. In the case of $E_f=-\frac{U}{2}$,  the Fermi energy of the conduction bands is zero which is just in the middle of the two energy levels of the $f$ orbital ($-\frac{U}{2}$ and $\frac{U}{2}$). Away from the axis of $E_f=-\frac{U}{2}$, considering the particle-hole symmetry, all the physical quantities map to each other.  From Fig. \ref{PD} we could see there are para-magnetic (PM) phase and two different anti-ferromagnetic phases (AF1 and AF2). The magnetic transition is shown as blue lines in Fig. \ref{PD}. From AF1 phase the magnetic transition is continuous, while from AF2 phase it's first order. In Fig. \ref{PD} the continuous magnetic transition is displayed as the blue dashed line, and the first order magnetic transition is the blue solid line. The phase transition between the two magnetic order is of first order, and displayed as red solid line in Fig. \ref{PD}. It is a Lifshitz transition which accompanies by the reconstruction of Fermi surface. We will discuss this in more detail later. Inside the AF2 phase, there's a Kondo region. In the Kondo region the occupation number of electrons on $f$ orbital $n_f$ is $1$, and so are the $n_c$ and $n_d$. It's worth mentioning that the term ``Kondo region" doesn't mean Kondo effect takes place, we just follow the nomination in literature \cite{Callaway1988_PRB38-2583--2595}.

\begin{figure}
\centering
\includegraphics[width=0.238\textwidth]{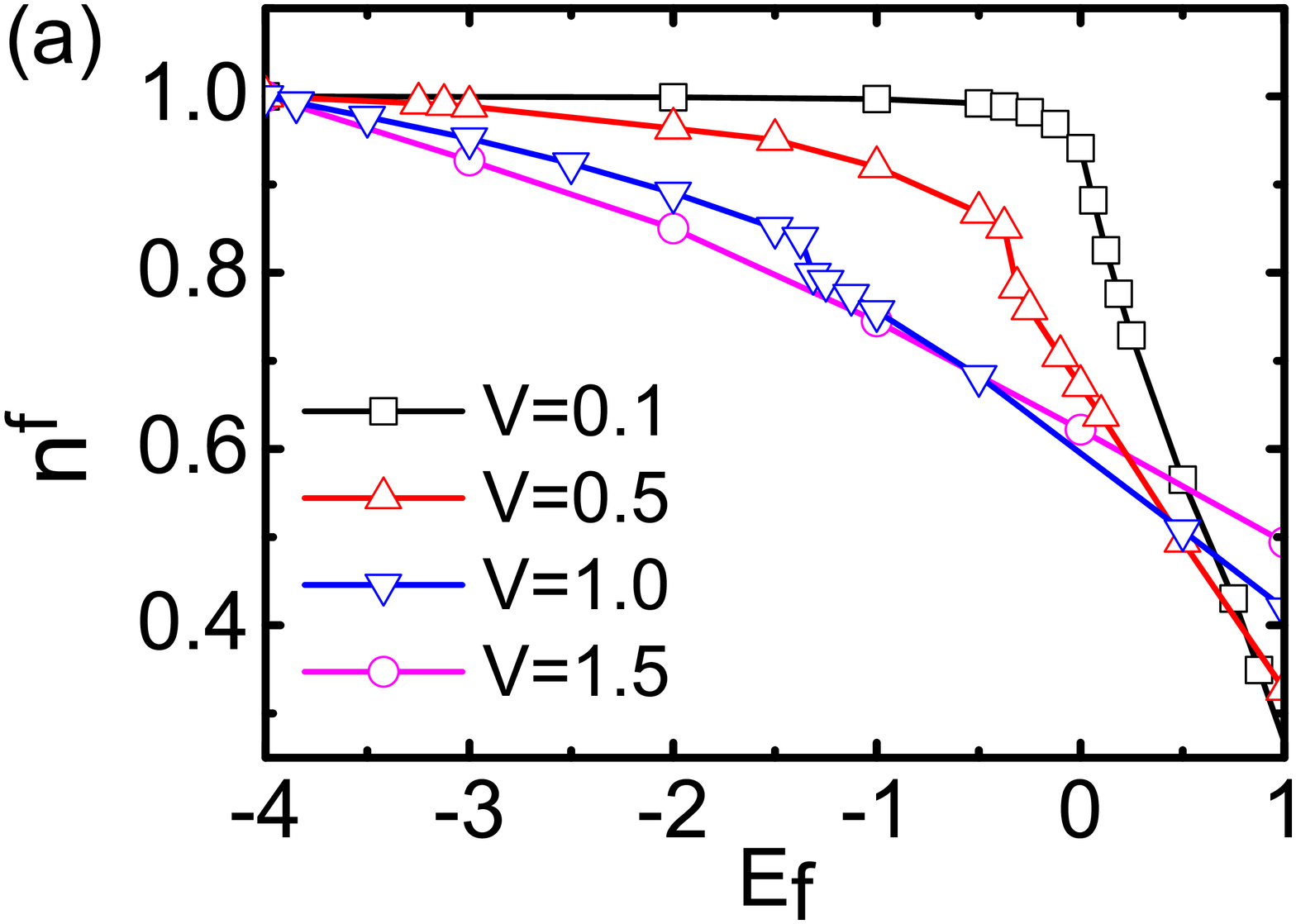}
\includegraphics[width=0.238\textwidth]{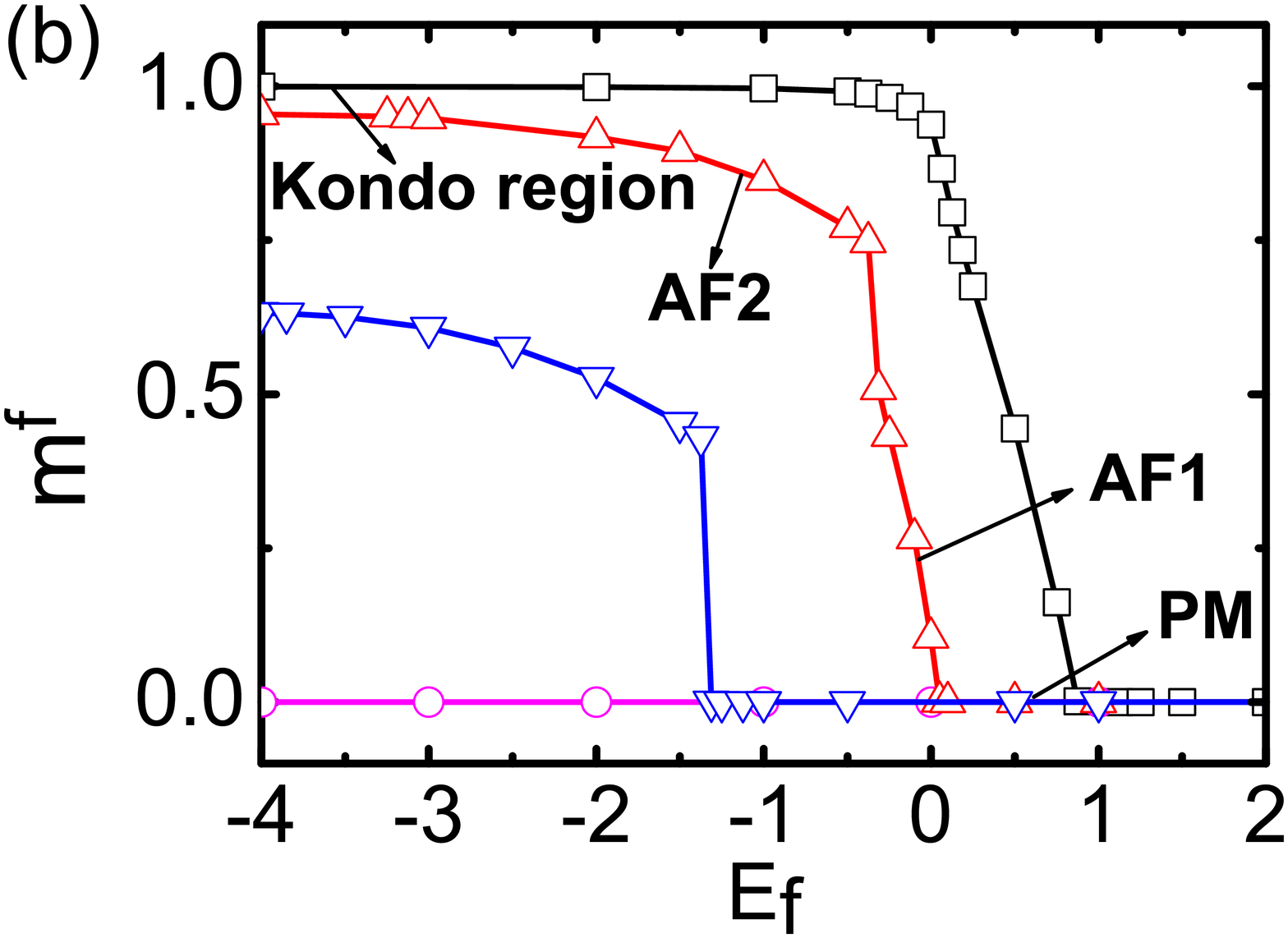}
\caption{(Color online) By varying $E_f$, the results when $V=0.1$, $V=0.5$, $V=1.0$ and $V=1.5$ are displayed with black, red, blue and magenta color. (a) The occupation number of $f$ orbitals. (b) The local magnetic moment of $f$ orbitals.
}
\label{nm}
\end{figure}

Now we discuss how the phase diagram is determined. We have calculated the local magnetic moments and the number of electrons, and they are shown in Fig. \ref{nm}. The definitions of those physical quantities are:
\begin{equation}
\begin{aligned}
m_i^\alpha&=\langle n_{i\uparrow}^\alpha - n_{i\downarrow}^\alpha \rangle
\\n_i^\alpha&=\langle n_{i\uparrow}^\alpha + n_{i\downarrow}^\alpha \rangle
\end{aligned}
\end{equation}
Here $n_{i\sigma}^\alpha$ is the number of spin $\sigma$ electrons of $\alpha$ orbitals on site $i$. As we mentioned before that $m_i^f$ and $n_i^f$ are symmetric with respect to $E_f=-\frac{U}{2}$, due to the particle-hole symmetry. In order to display more details we only plotted the data when $E_f>-\frac{U}{2}$.
At half filling $n_i^c+n_i^d+n_i^f=3.0$, considering $n_i^c=n_i^d$ in the symmetric case, so only $n_i^f$ is plotted.

At small value of $V$, there are mainly five regions: (1) Maximally occupied $f$ states where $n_f=2$, when $E_f<E_c^{(0)}-U$ ; (2) First mixed valence region where $1<n_f<2$, when $E_c^{(0)}-U<E_f<E_c^{(1)}-U$ ; (3) Kondo region where $n_f=1$, when $E_c^{(1)}-U<E_f<E_c^{(1)}$ ; (4) Second mixed valence region where $0<n_f<1$, when $E_c^{(1)}<E_f<E_c^{2}$ ; (5) Empty $f$ states where $n_f=0$, when $E_f>E_c^{(2)}$. The $E_c^{(0)}$ ($E_c^{(2)}$) is the lowest (highest) energy level of the conduction band, and $E_c^{(1)}$ is the fermi energy of the conduction band. In Fig. \ref{nm}(a) only the Kondo region ($-8<E_f<0$) and the second mixed valence region ($0<E_f<4$) are shown. As the hybridization strength $V$ increases the two mixed valence regions expands, at the same time the other three regions shrink. The Kondo region becomes smaller and smaller as $V$ increases, and for $V\approx1.20$ it becomes a point and only the symmetric point $E_f=-4$ belongs to the Kondo region. At the symmetric point $E_f=-4$, $n^f=1$ no matter how the hybridization strength $V$ changes.

The anti-ferromagnetic long range order is formed in the Kondo region when the value of $V$ is small. If $V$ is fixed, and $E_f$ goes away from the symmetric axis, the magnetic transition to a para-magnetic phase takes place in the mixed valence region. It's obvious in Fig. \ref{nm} that there exists a sizable jump in both local magnetic moment and number of electrons when $V=0.5$ and $V=1.0$ (the red and blue line in Fig. \ref{nm}). This is due to the occurrence of Lifshitz transition. Even though there has been several studies of Lifshitz transition on PAM\cite{Kubo2013_PRB87-195127, Wysoki2014_PRB90081114, KuboKatsunori2015_JPSJ84-094702}, it happening at half filling is still unusual \cite{Jian-WeiYang2018_ChinPB27-37101}.

\begin{figure}
\includegraphics[width=0.45\textwidth]{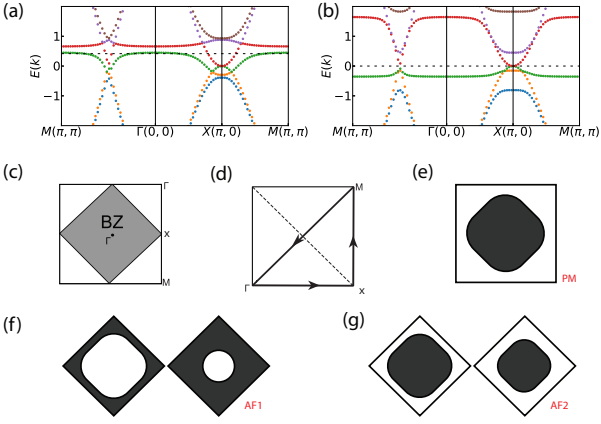}
\centering
\caption{(Color online)  (a) Band structure in the AF1 phase ;(b) Band structure in the AF2 phase ; The dashed line is the Fermi level at half filling ; (c) The first Brillouin zone of the square lattice, and the grey shaded region is the first Brillouin zone when anti-ferromagnetic order is present; (d) The upper right quarter of the Brillouin zone. The band structure are plotted from $M$ to $\Gamma$, $\Gamma$ to $X$, and $X$ to $M$ as the arrows indicated; (e) The Fermi surface in the PM phase ; (f) The Fermi surface in the AF1 phase ; (g) The Fermi surface in the AF2 phase(slightly away from half filling).
}
\label{AFdispersion}
\end{figure}

In order to understand how the Lifshitz transition occurs, we plotted the band structure of the two anti-ferromagnetic phases in Fig. \ref{AFdispersion}. As we mentioned in the previous section, in a DMET calculation, the correlation potential is self-consistently determined. Adding a converged correlation potential to the non-interacting part of the Hamiltonian, and diagonalizing the auxiliary Hamiltonian, the band structure could be derived. The presence of anti-ferromagnetic order makes the unit-cell twice than before, so the first Brillouin zone of the reciprocal lattice becomes half of the non-magnetic case. The bigger square in Fig. \ref{AFdispersion}(c) is the first Brillouin zone ($\Gamma$ point is in the center) in the PM phase, and the grey shaded smaller square is the first Brillouin zone in the presence of anti-ferromagnetic order. Fig. \ref{AFdispersion}(d) is a quarter of the upper panel with all the high symmetry point marked. From Fig. \ref{AFdispersion}(c) we know the Brillouin zone is folded along two neighboring $X$ point. The band structure repeats itself along the dashed line of Fig. \ref{AFdispersion}c. So there are six bands in the band dispersion figures of the two AF phases. Fig. \ref{AFdispersion}(a)(b) are the band structure of AF1 phase and AF2 phase, and the Fermi level at half filling are displayed as dashed line. The AF1 phase has a hole type Fermi surface around $\Gamma$ point. The topology of the AF1 phase is the same as PM phase. However the AF2 phase is rather different. At half filling, it's in a semi-metal phase, since X point and the middle point between $\Gamma$ point and M point have "Dirac cone". Please note the Fermi surface in Fig. \ref{AFdispersion}(g) is the Fermi surface slightly away from half filling. The Fermi surface of AF1 phase and AF2 phase is similar with the previous Kondo lattice model studies \cite{Watanabe2007_PRL99-136401,Peters2015_PRB92-075103}. The AF1 phase has a hole-type large Fermi surface, and the AF2 phase has an electron-type small Fermi surface (at half filling AF2 phase is in a semi-metal phase, and there's only "Fermi line"). The difference of the three orbital model and the standard two band model is that two band are crossing the Fermi energy level instead of one band. 

\subsection{Spin correlations and Kondo singlet}
\begin{figure}
\centering
\includegraphics[width=0.238\textwidth]{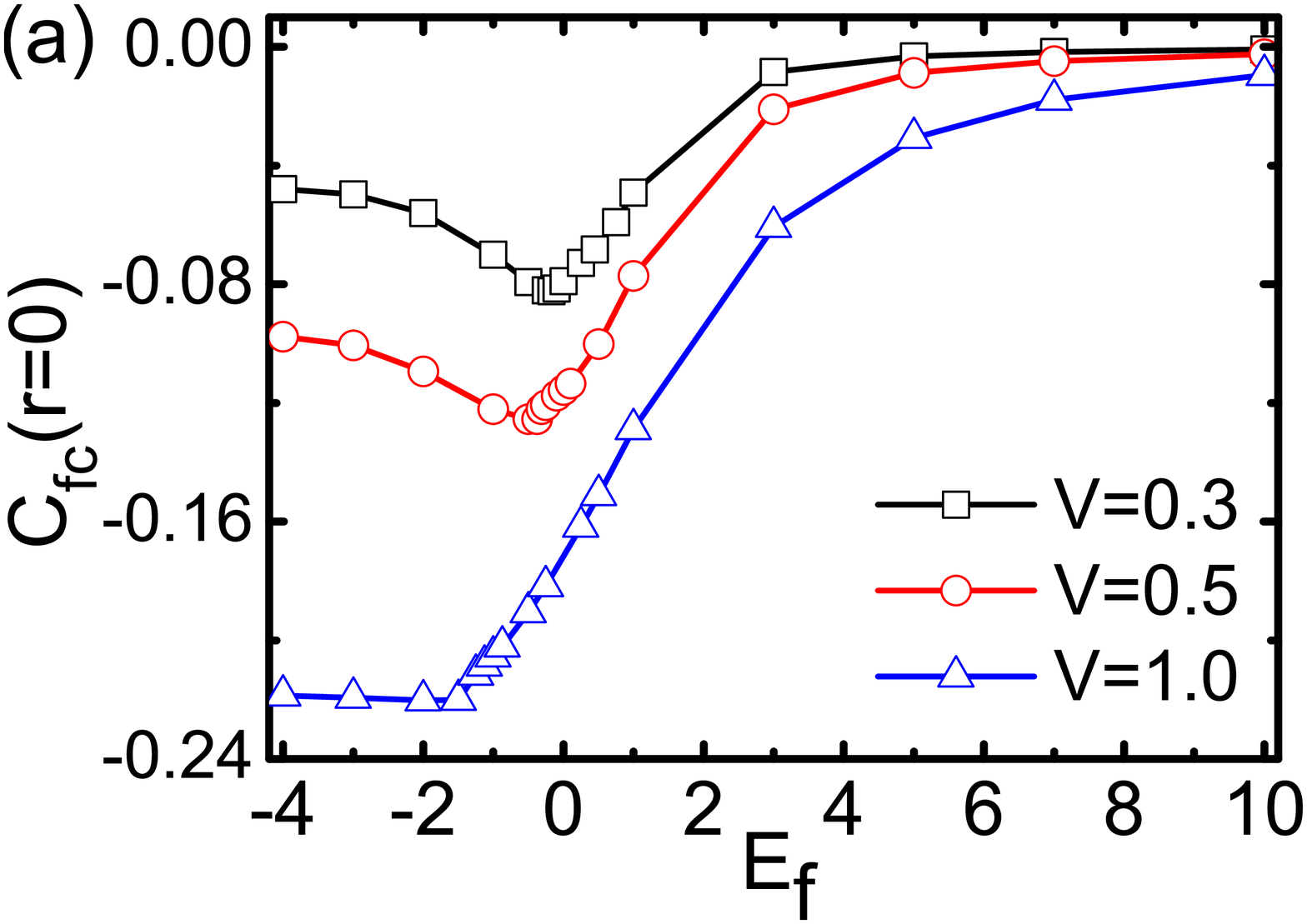}
\includegraphics[width=0.238\textwidth]{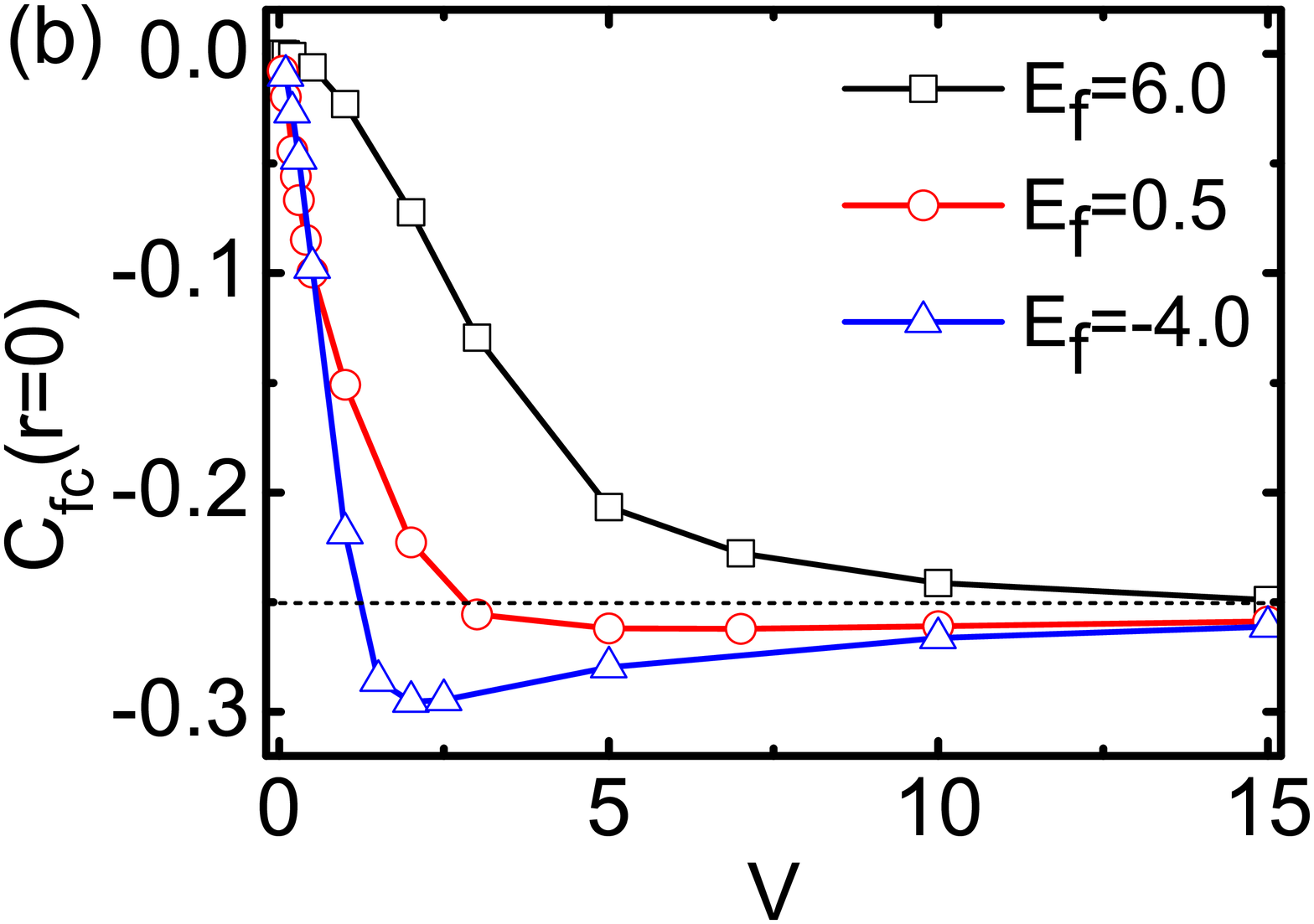}
\includegraphics[width=0.238\textwidth]{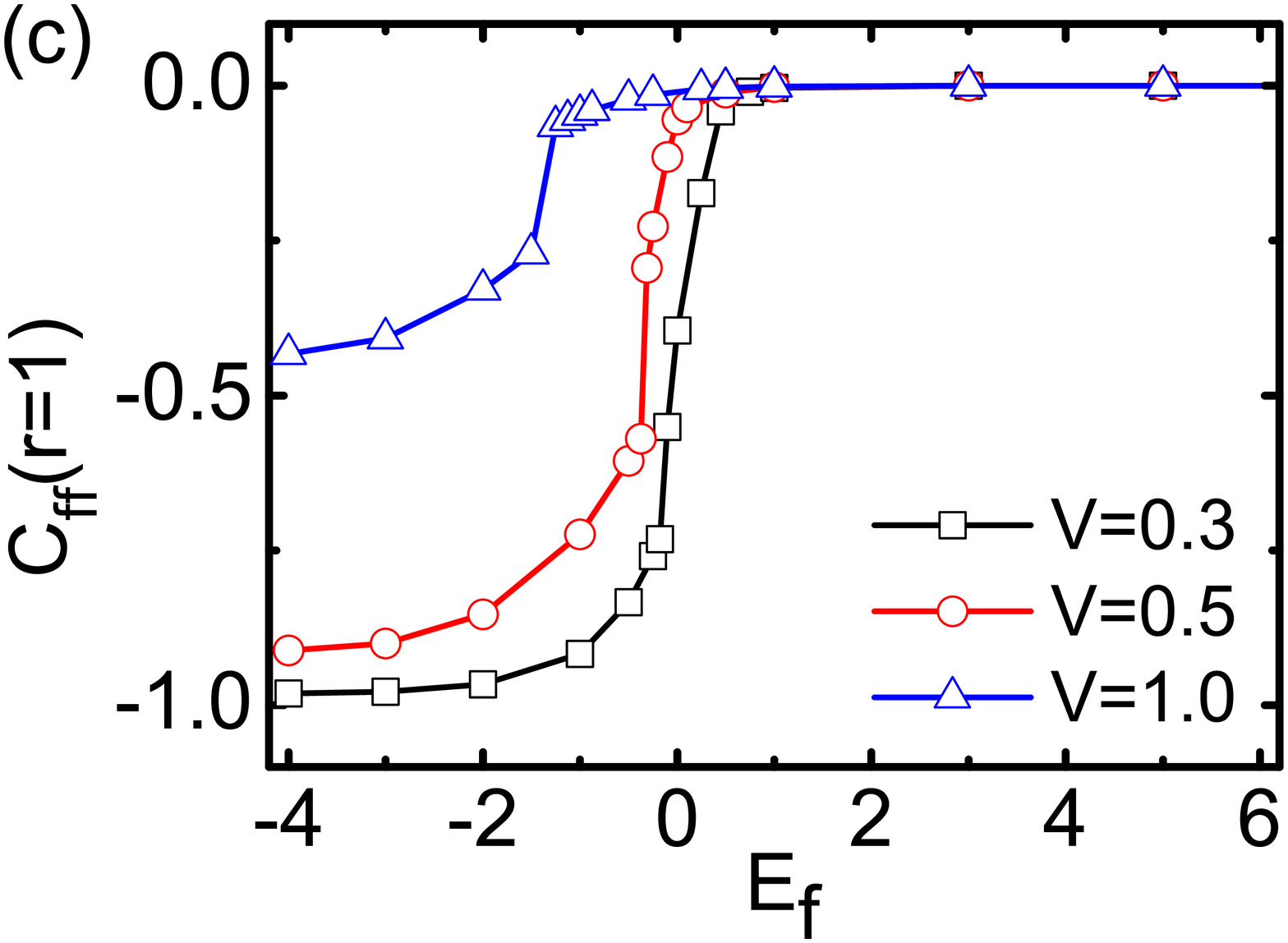}
\includegraphics[width=0.238\textwidth]{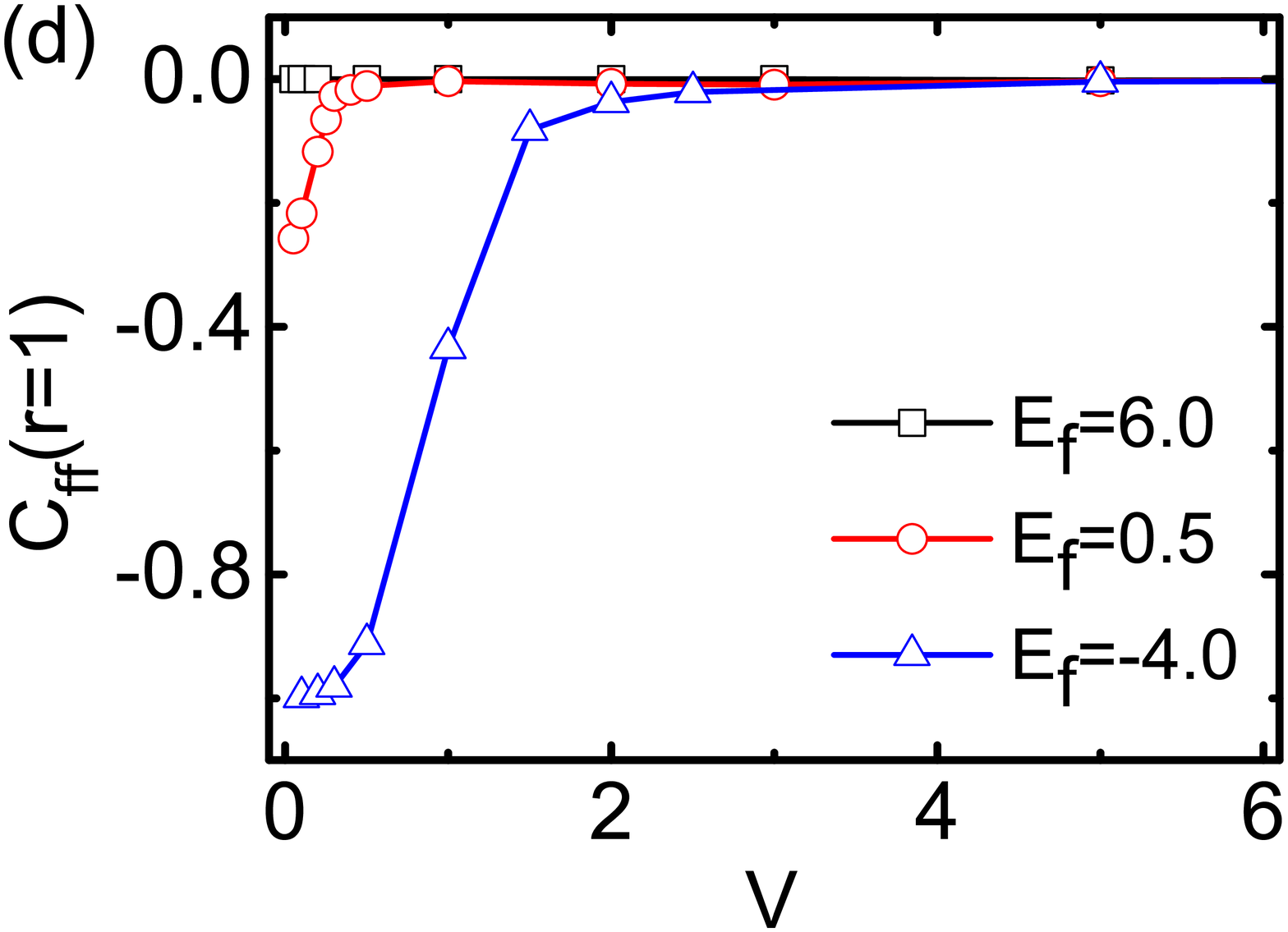}
\caption{(Color online) The spin-spin correlation functions as a function of $E_f$ in left panels, and as a function of $V$ in the right panels. (a) and (b) are the magnetic correlation between intrasite $f$ orbital and $c$ orbital. (c) and (d) are the magnetic correlation between neighboring $f$ orbitals.
}
\label{Correlation-Ef}
\end{figure}

The magnetic physics of the PAM can be characterized by the spin-spin correlations. We study the spatial spin-spin correlation functions, and the definitions are:
\begin{equation}
\begin{aligned}
C_{fc}(r=0)&=\langle m_i^f \cdot m_i^c \rangle=
\langle (n_{i\uparrow}^f-n_{i\downarrow}^f)(n_{i\uparrow}^c-n_{i\downarrow}^c)\rangle\\
C_{ff}(r=1)&=\langle m_i^f \cdot m_j^f \rangle=
\langle (n_{i\uparrow}^f-n_{i\downarrow}^f)(n_{j\uparrow}^f-n_{j\downarrow}^f)\rangle
\end{aligned}
\end{equation}
here $C_{fc}(r=0)$ measures the magnetic correlations between the localized orbital $f$ and conduction orbital $c$ on the same site. While $C_{ff}(r=1)$ measures the correlations of localized orbital $f$ between neighboring sites. To explore the magnetic property the hybridization $V$ is fixed first. The results are displayed in the left panel of Fig. \ref{Correlation-Ef}. Starting from the symmetric point at $E_f=-4$ the system evolves from AF2 phase to PM phase directly when $V=1.0$. In the AF2 phase the correlation function $C_{fc}(r=0)$ is almost constant, and it drops to zero gradually in the PM phase. Moreover there's a kink at the magnetic transition point. While the behavior of $C_{ff}(r=1)$ is rather similar to the local magnetic moment. Its absolute value decreases slowly in the AF2 phase, and after a finite step, it approaches to zero gradually. Apart from the discontinuous of $n^f$ and $m^f$, the jump here is another evidence that the transition from AF2 phase to PM phase is first order. Meanwhile the absolute value of $C_{fc}(r=0)$ is smaller when $V=0.3$ and $V=0.5$. The reason is the hybridization strength $V$ increases the anti-ferromagnetic spin-spin interaction between $f$ and $c$ orbitals. The system undergoes all the three phases when $V=0.3$ and $V=0.5$. The absolute value of $C_{fc}(r=0)$ increases slightly in the AF2 phase. It mainly decreases in the AF1 phase, and of course becomes to zero eventually in the PM phase. However the minimum point of the $C_{fc}(r=0)$ curve is not the Lifshitz transition point. Unlike the order parameter there's no sudden change when entering in a new phase.

Next we check the results in the right panel of Fig. \ref{Correlation-Ef}. They are calculated by fixing $E_f$ and varying $V$ continuously. Different colors in the figures represent different value of $E_f$. If the value of $V$ is small, $E_f=-4$, $E_f=0.5$ and $E_f=6$ correspond to the Kondo region, the second mixed valence region and empty $f$ states. The behavior of $C_{ff}(r=1)$ is easy to interpret. When $E_f=6$ the anti-ferromagnetic long range order never shows up, so it keeps to zero at all value of $V$. For the other two values of $E_f$, the long range order is present at small value of $V$, so there's anti-ferromagnetic correlations between neighboring $f$ orbitals. As $V$ increases, it becomes to zero gradually.
However the $C_{fc}(r=0)$ curves are more interesting. Although there's no anti-ferromagnetic long range order when $E_f=6$, the anti-ferromagnetic correlation between $f$ and $c$ orbitals increases monotonously as $V$ increases. While the situation when $E_f=0.5$ and $E_f=-4$ is different. If the long range order presents, the absolute value of $C_{fc}(r=0)$ increases rapidly. And in the PM phase it increases slightly and then decreases very slowly. At large value of $V$, regardless of the value of $E_f$, the value of $C_{fc}(r=0)$ approaches $-1/4$ (dashed line in Fig. \ref{Correlation-Ef}(b)). This suggests that the paramagnetic phase at large value of $V$ is different from the phase when $E_f$ is fay away from the symmetric point.

The hybridization between conduction and localized orbitals is responsible for the creation of the Kondo singlet, and in the mean field level the hybridization parameters are introduced to qualify the formation of Kondo singlet \cite{Asadzadeh2013_PRB87-205144,Li2015_JPCM27-425601}. In order to characterize the Kondo screening, a hybridization parameter $V_u$ is defined as:

\begin{equation}
V_u = -\frac{1}{2} (V_a+V_b)
\label{hybridization}
\end{equation}

here $V_a$ and $V_b$ are hybridization parameters defined on the two sublattices A and B:  :
\begin{equation}
\begin{aligned}
V_a&=\langle c^\dag_{iA\uparrow} f_{iA\uparrow} \rangle &= \langle c^\dag_{iB\downarrow} f_{iB\downarrow} \rangle \\
V_b&=\langle c^\dag_{iA\downarrow} f_{iA\downarrow} \rangle &= \langle c^\dag_{iB\uparrow} f_{iB\uparrow} \rangle
\end{aligned}
\end{equation}

\begin{figure}
\centering
\includegraphics[width=0.238\textwidth]{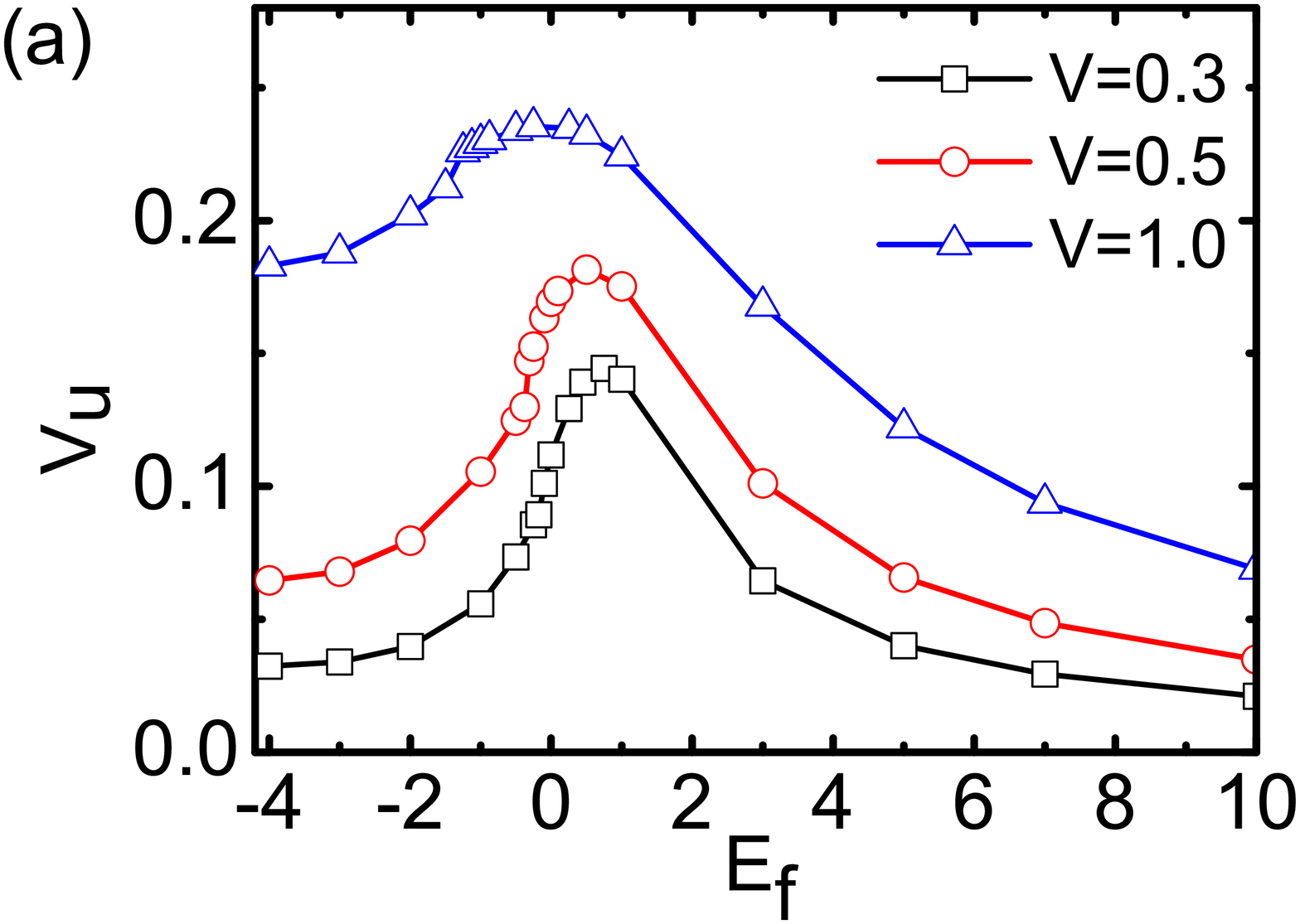}
\includegraphics[width=0.238\textwidth]{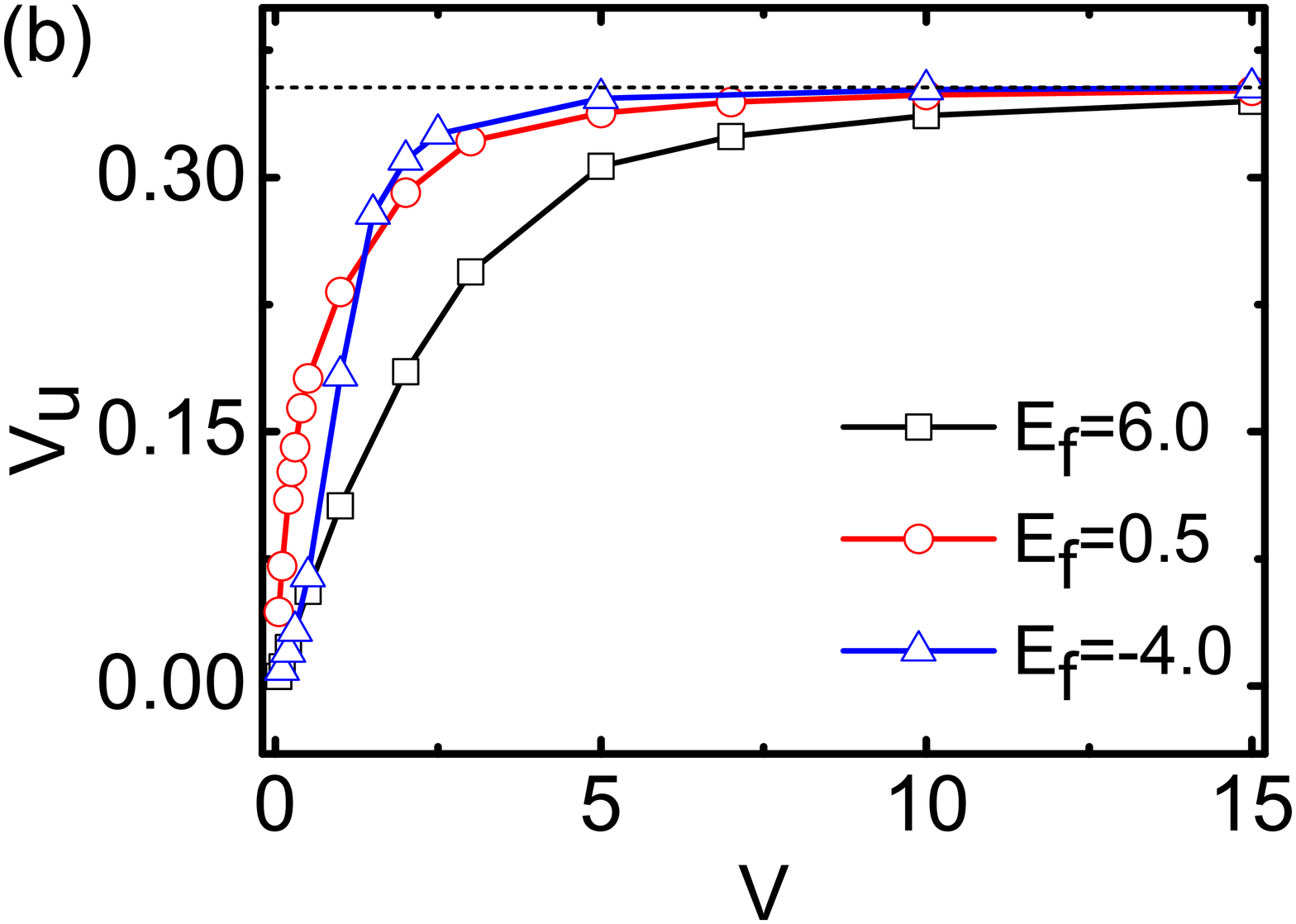}
\caption{(Color online) The hybridization parameter $V_u$ defined in Eq. \ref{hybridization} to characterize the Kondo screening. (a) $V_u$ as a function of $E_f$. (b) $V_u$ as a function of $V$, the dashed line is $1/3\approx0.3333$.
}
\label{Vu}
\end{figure}

Now we discuss how the hybridization parameter $V_u$ varies with different parameters, the results are shown in Fig. \ref{Vu}. Since the Kondo coupling $J$ between the localized moment and conduction electrons is proportional to the square of hybridization strength $V$, it's more and more likely to form Kondo singlet as $V$ increases. The three different curves in the left panel of Fig. \ref{Vu} accord with the fact that Kondo effect dominates more as $V$ increases. In both the AF1 phase and AF2 phase the hybridization parameter $V_u$ increases as $E_f$ increases. This indicates the competition between the Kondo effect and RKKY effect, as the RKKY effect becomes weak as $E_f$ is away from the symmetric point. However in the PM phase $V_u$ decreases as $E_f$ goes away from the symmetric point. This is due to the number of electrons in $f$ orbitals are descending as $E_f$ increases. Near the magnetic transition point $V_u$ reaches its maximum. If AF1 phase is present, the maximum is located in the AF1 phase, otherwise the maximum is in the PM phase.

By fixing $E_f$ the hybridization parameter $V_u$ increases as $V$ increases monotonously. At large value of $V$, the  hybridization parameter $V_u$ approaches $0.5$. In both the AF1 phase ($E_f=0.5$, $V<0.285$) and AF2 phase ($E_f=-4$, $V<1.325$) the hybridization parameter $V_u$ increases rapidly. After entering into the PM phase, the slope of $V_u-V$ becomes smaller and smaller. When the hybridization strength $V$ is large enough $V_u$ approaches $1/3\approx0.3333$, that is plotted as dashed line in Fig. \ref{Vu}(b).

\subsection{Non-symmetric case $V_1 \neq V_2$}
We further consider the non-symmetric case when $V_1 \neq V_2$. The hybridization strength $V_1=V$ and $V_2=\gamma V_1$,
and in the following only $V$ is varied continuously. It's surprising that the data with different hybridization ratio $\gamma$ are all connected with each other through a ``scaling transformation". After the transformation all the data collapse just as the finite size scaling.

\begin{figure}
\centering
\includegraphics[width=0.45\textwidth]{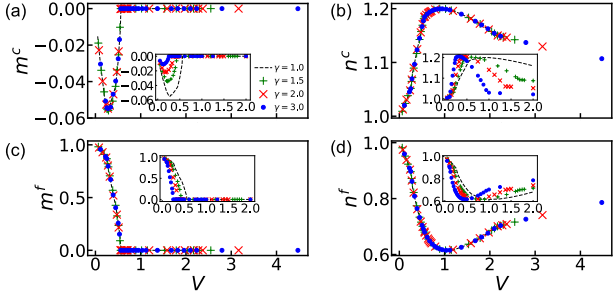}
\caption{(Color online) With the transformation in Eq. \ref{transform1}-\ref{transform2}, the data of $m^{\alpha}$ and $n^{\alpha}$ when $E_f=-1$, $V_1=V$ and $V_2=\gamma V$. The insets are the data before the transformations. (a) local magnetic moment of $c$ orbital. (b) occupation number of electrons on $c$ orbital. (c) local magnetic moment of $f$ orbital. (d) occupation number of electrons on $f$ orbital.
}
\label{11}
\end{figure}

The scaling transformation of local magnetic moment is:
\begin{equation}
\begin{aligned}
m^c \left(V,\gamma V\right) &= \frac{2}{1+\gamma^2}\bar{m}^c\left(\sqrt{\frac{1+\gamma^2}{2}}V\right)\\
m^d \left(V,\gamma V\right) &= \frac{2\gamma^2}{1+\gamma^2}\bar{m}^c\left(\sqrt{\frac{1+\gamma^2}{2}}V\right)\\
m^f \left(V,\gamma V\right) &= \bar{m}^f\left(\sqrt{\frac{1+\gamma^2}{2}}V\right)
\end{aligned}
\label{transform1}
\end{equation}
here $\bar{m}^{\alpha}$ is the local magnetic moment on orbital $\alpha$ in the case of $V_1=V_2=V_0$, and the value of the hybridization strength $V_0$ is $\sqrt{\frac{1+\gamma^2}{2}}V$. After the transformation  $m^c$ and $m^f$ are displayed in Fig. \ref{11}(a)(c), and the insets are the original data from simulation.  Different ratio of the hybridization strength are displayed with different colors, $\gamma=1$,$\gamma=1.5$,$\gamma=2$, and $\gamma=3$ is black, green, red, and blue in Fig. \ref{11}. In this model the only difference between orbital $c$ and orbital $d$ is the hybridization strength. $m^d$ is equivalent with $m^c$ through $\gamma\rightarrow1/\gamma$, thus only $m^c$ is displayed.

However the formula of the transformation for $n^c$ and $n^d$ is a bit more complex :
\begin{equation}
\begin{aligned}
n^c \left(V,\gamma V\right) &= \frac{2}{1+\gamma^2}\bar{n}^c\left(\sqrt{\frac{1+\gamma^2}{2}}V\right)+\frac{\gamma^2-1}{\gamma^2+1}n^g\left(V,\gamma V\right)\\
n^d \left(V,\gamma V\right) &= \frac{2\gamma^2}{1+\gamma^2}\bar{n}^c\left(\sqrt{\frac{1+\gamma^2}
{2}}V\right)-\frac{\gamma^2-1}{\gamma^2+1}n^g\left(V,\gamma V\right)\\
n^f \left(V,\gamma V\right) &= \bar{n}^f\left(\sqrt{\frac{1+\gamma^2}{2}}V\right)
\end{aligned}
\label{transform2}
\end{equation}
As in Fig. \ref{11}(b)(d), after the transformation all the data collapses. This behavior is quite similar to the finite size scaling. Since the equivalence of orbital $c$ and orbital $d$, as well as $n^c+n^d+n^f=3.0$, only the results of $n^c$ and $n^f$ are displayed.

\begin{figure}
\centering
\includegraphics[width=0.45\textwidth]{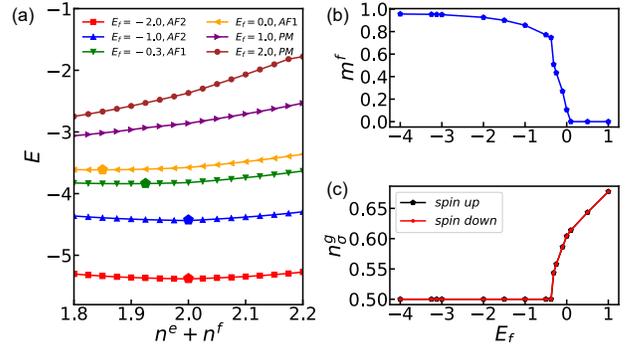}
\caption{(Color online) In the standard PAM $V'=1/\sqrt{2}$ such that in the corresponding three orbital model $V_1=V_2=0.5$.  (a) The ground state energy of the standard PAM as a function of occupation number, and the lowest energy is marked with pentagons. (b) The $f$ orbital local magnetic moment as a function of $E_f$. (c)The number of electrons on $g$ orbtial as a fucntion of $E_f$.
}
\label{13}
\end{figure}

In order to understand the physical meaning of $n^g$, now we consider the extreme case of $V_1=0$ and $V_2=V'\equiv\sqrt{1+\gamma^2}V_1$. The $d$ orbital and $f$ orbital constitute a standard PAM, and $c$ orbital is alone. To distinguish from previous paragraphs, $c$ orbital is mentioned as $g$ orbital, and $d$ orbital  is mentioned as $e$ orbital. At half filling still $n^e+n^f+n^g=3.0$. We simulate the system constituted by standard PAM and one separated conduction band, and keep the occupation number of electrons to half filling, the results are presented in Fig. \ref{13}. The total ground state energy versus the number of $n^e + n^f$ is shown in Fig. \ref{13}(a). Different colors represent different $E_f$. The value of $V'$ is chosen that the equivalent $V_1 = V_2 = 0.5$. In the AF2 phase the ground state of the system constitutes  $e$ and $f$ orbitals at half filling. In the AF1 and PM phase as $E_f$ increase, the standard PAM is less than half filling, and the electrons become more likely to stay in the $g$ orbital. Fig. \ref{13}(b) is the local magnetic moment on $f$ orbital, and Fig. \ref{13}(c) is the occupation number of the $g$ orbital. All the results here are consistent with each other, and they suggested how the Lifshitz transition occurs at half filling. In the standard PAM the Fermi surface reconstruction happens away from half filling \cite{Jian-WeiYang2018_ChinPB27-37101}.

\begin{figure}
\centering
\includegraphics[width=0.45\textwidth]{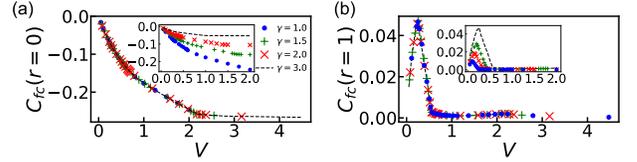}
\caption{(Color online) The spin-spin correlation functions when $E_f=-1$, $V_1=V$ and $V_2=\gamma V$. The data before the transformations in Eq. \ref{transform3} are displayed in the insets. (a) magnetic correlation function between intrasite $c$ and $f$ orbitals. (b) magnetic correlation function between $c$ and $f$
orbitals on neighboring sites.
}
\label{12}
\end{figure}

In a word the non-symmetric case is related with the symmetric case as long as $V_1^2+V_2^2=V_0^2$. Both $m^{\alpha}$ and $n^{\alpha}$ are two fermionic operators, and the above transformations are held for all the data with different ratio of hybridization strength. Now we consider the four fermionic operators, such as the spin-spin correlation functions. Still the transformations exist: 
\begin{equation}
\begin{aligned}
C_{fc}(r=0)=\langle m_i^f \cdot m_i^c \rangle&=\bar{C}_{fc}(r=0)=\frac{2}{1+\gamma^2}\langle \bar{m}_i^{f} \cdot \bar{m}_i^{c} \rangle\\
C_{fc}(r=1)=\langle m_i^f \cdot m_i^c \rangle&=\bar{C}_{fc}(r=1)=\frac{2}{1+\gamma^2}\langle \bar{m}_i^{f} \cdot \bar{m}_j^{c} \rangle\\
C_{fd}(r=1)=\langle m_i^f \cdot m_i^d \rangle&=\bar{C}_{fd}(r=1)=\frac{2\gamma^2}{1+\gamma^2}\langle \bar{m}_i^{f} \cdot \bar{d}_j^{d} \rangle\\
C_{ff}(r=1)=\langle m_i^f \cdot m_j^f \rangle&=\bar{C}_{ff}(r=1)=\langle \bar{m}_i^{f} \cdot \bar{m}_j^{f} \rangle\\
\end{aligned}
\label{transform3}
\end{equation}
The notations are the same as previous, and the results are displayed in Fig. \ref{12}. We don't want to bother the readers with all the data, so only $C_{fc}(r=0)$ and $C_{fc}(r=1)$ are displayed. 

\section{Conclusions}
A number of theoretical and numerical work \cite{Potthoff1999_PRB59-2549--2555,Potthoff1999_PRB60-7834--7849,Okamoto2004_PRB70-241104,
Ishida2012_PRB85-045112,Helmes2008_PRL101-066802,Euverte2012_PRL108-246401} have examined the physics at the interface of Mott insulator and metal. Inhomogeneous DMFT predicts fragile fermi liquid appears in finite layers of Mott insulator sandwiched between metallic leads \cite{Zenia2009_PRL103-116402}. In this paper, by introducing a three orbital periodic Anderson model, we have studied one insulator layer sandwiched between two metallic layers with DMET.

The model we studied is a periodic Anderson model with degenerate conduction orbitals. We start with the symmetric case, when the two conduction orbitals have equal hybridization strength with the localized orbital. We found there are three different phases at half filling. When the hybridization strength $V$ is weak, the RKKY effect dominates, and the ground state is in anti-ferromagnetic phase. As $V$ increases the Kondo effect becomes important, and para-magnetic phase appears. In the region when $V$ is small, there exists two different anti-ferromagnetic phases.

The phase transition between the two AF phases is the Lifshitz transition, which is accompanied by the Fermi surface reconstruction. From the band structure, we discussed the topology of the Fermi surface. We further studied the non-symmetric case, and found the equivalence of the model to another model. In the picture of the other model, the mechanism of the Lifshitz transition is more clear. We also studied the spin-spin correlation functions carefully. When $V$ is small, even though the Kondo effect is not that strong, as $E_f$ is away from the symmetric point, the Kondo effect becomes more important at first, then disappears as expected. However the quantization of the strength of Kondo screening is not well defined, otherwise it will be interesting to unearth it deeply.

Compared with the standard PAM, the phase diagram of the thee orbital PAM is more rich at half filling. There is only one anti-ferromagnetic phase at half filling in the standard PAM \cite{Jian-WeiYang2018_ChinPB27-37101}. While there are two different anti-ferromagnetic phases in the three orbital model. If $V<0.75$ both AF1 phase and AF2 phase are present, and if $V>0.75$ AF1 phase is absent. The Fermi surface is also different from the standard model. Two bands are crossing the Fermi level in the three band model. Further more AF2 phase is in a semi-metal phase at half filling. Away from half filling, AF2 phase enters into the metal phase. From the band structure, the phase diagram of the three orbital model would be more complex away from half filling. Although with so many differences, the three orbital model has connections with the standard PAM. It's equivalent with the standard PAM along with a non-interacting band.


Our work on the three orbital PAM is a first step in the applications of DMET to superlattice $f$ electron models. We only restricted ourselves at half filling. There will be more exotic and fascinating phenomena far away from half filling, such as complex magnetic order, unconventional superconductivity, and exotic transport properties. The model we studied here is too simple to describe any real materials. The extra correlated layer drives the system into the semi-metal phase. But the semi-metal phase only appears at half filling. It's difficult to predict any observable effects in experiments, with only static zero temperature physical properties. The transport properties and thermodynamics would be interseting, and they will be the next step. Our results indicate the physics of the quasi-two dimensional model is different from the standard model's. In order to study more complex and realistic system, developing more powerful impurity solvers, with high precision and low computational cost will be significant. 


\begin{acknowledgments}
This work is supported by the National Science Foundation of China (Grant Nos. 11504023 and 11374034), and Beijing Science Foundation (Grant No. 1192011). We are grateful for the fruiteful discussions with Tao Li. We thank Yin Zhong for useful comments on the manuscript, and Boxiao Zheng for his help to overcome the convergence problem. We acknowledge National Super Computer Center in Tianjin for computing time.
\end{acknowledgments}

\bibliographystyle{apsrev4-1}
\bibliography{squarePAM.bib}

\end{document}